\documentclass[conference]{IEEEtran}
\ifCLASSINFOpdf
\usepackage[pdftex]{graphicx}
\graphicspath{{figure/}{../jpeg/}}
\DeclareGraphicsExtensions{.pdf,.jpeg,.png}
\else
\usepackage[dvips]{graphicx}
\graphicspath{{../figure/}}
\DeclareGraphicsExtensions{.eps}
\fi
\begin{document}

	%
	\title{A readout method based on 10 Gigabit Ethernet\\  for Sipixel detector}

	\author{\IEEEauthorblockN{Hangxu Li, Jie Zhang, Jingzi Gu, Xiaolu Ji, Yang Li, Jun Hu,  Yunhua Sun, Xiaoshan Jiang, Zheng Wang}}

	\maketitle
	
	\begin{abstract}
		With the rapid development of network protocol, TCP/IP has been widely applied in various OS systems because of its advantages of high-speed network transmission and standardization. Furthermore, TCP/IP has the functions of retransmission and flow control, so it can ensure data's security and stability. In recent years, as a transmission Field-bus, it is also widely used in high-energy physics detectors. This article describes a readout method using 10 gigabit network processor applied in silicon pixel detector which is a pixel detector designed for the High Energy Photon Source (HEPS) in China. The PHY layer and the MAC layer of this hardware are both instantiated in FPGA. The logic of hardware can process and read out the data of the detectors. Meanwhile, it wraps up data and send them to TOE (TCP Offload Engine). A kind of bus control protocol interface is also provided, which can facilitate the host computer to read and write the specific register directly through the UDP protocol. Only a piece of FPGA can complete the processing of the TCP and UDP packets. It is important for the miniaturization of the detectors.
	\end{abstract}
	

	%
	\IEEEpeerreviewmaketitle

	\section{Introduction}
	HEPS is a read-out electronics system based on the BPIX\cite{wei2015high} pixel array detector readout chip which is independently designed and developed by the Institute of High Energy Physics. The read-out electronics system uses Field-Programmable Gate Array (FPGA) as the core of digital logic. In order to complete the basic function of the pixel detector, the whole system need to be set up the configuration and read out data chain of the BPIX read-out chips. The prototype of the pixel detector consists of 16 sensor modules, covering an effective detection area of 17.28 cm * 12.48 cm. The detection energy interval is 8 keV -20 keV\cite{gu2017high}.By way of transmitting data in real-time at 1K refresh rate, each bandwidth of read-out chain needs at least 4.8Gbps. 
	
	%

	\section{Hardware and firmware implementation}	
	\subsection{hardware implementation}
	
	FPGA is the logic core of the data readout firmware. The readout structure of a single module is shown in Figure 1 and 2. The FPGA's IO is directly connected to the pin of the BPIX readout chip so the serial data is transmitted to FPGA. At the same time, the hardware uses DDR3 as data cache and then processes and transfers packets through the high-speed GTX transceiver\cite{backplanesieee}. For single-module readout system, PHY layer connects directly to the SFP plus interface and through the optical fiber to the upper computer\cite{zhang2017heps}.
	
	\begin{figure}[htbp]
		\centering
		\includegraphics[width=6.5cm]{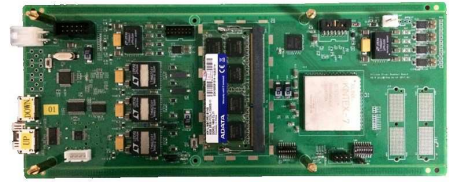}
		\caption{read-out board}
		\label{fig_sim}
	\end{figure}
	
	\begin{figure}[htbp]
		\centering
		\includegraphics[width=6.1cm]{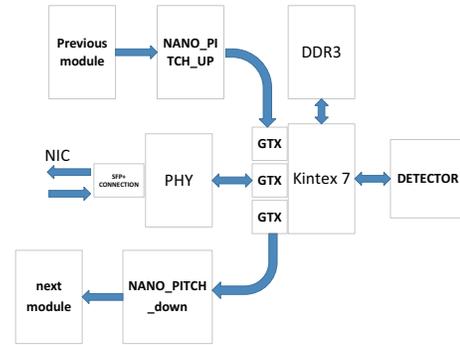}
		\caption{hardware structure}
		\label{fig_sim}
	\end{figure}
	
	\subsubsection{XTOE}
	
	The 10G TOE and 10 G EMACs are standard IP blocks that can be an interface between User logic via FIFO input or output buses and an XGMII interface of a 10Gb Ethernet PHY. The Full TCP Offload Engine implements standard TCP/IP transfer protocol with most widely used TCP options, over 10Gb Ethernet. The payload data is sent or received from user hardware via the user FIFO interfaces.
	
	\subsubsection{firmware structure}	
	
	The basic firmware framework is shown in Figure 3. Fast Control module receives external clock and completes clock processing. It also generates the 80Mhz clock required by other internal logic. Simultaneously, this module forwards trigger and completes the function of the signal number decoding. The REG Control module manages the internal registers which can be read or wrote through UDP parsing module. The BPIX module utilizes the registers' information to configure the parameter of ASIC and read data out from the sensors at the same time. As the most important part of the firmware, XTOE module plays the role of TCP/UDP data transmission. It also applies the mode of daisy chain to connect other read-out boards.
	\begin{figure}[htbp]
		\centering
		\includegraphics[width=6.5cm]{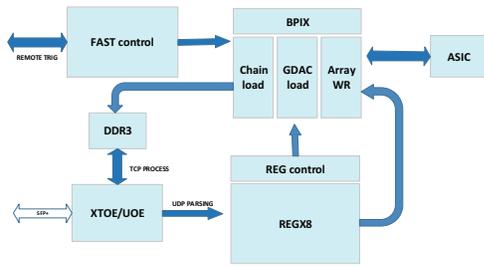}
		\caption{firmware structure}
		\label{fig_sim}
	\end{figure}
	
	The XTOE module is mainly divided into two parts, one is the TCP data processing module and the other is the UDP data parsing module. The way of TCP/UDP data transmission between XTOE and user layer is strictly adhered to the timing relationship required by XTOE IP core.
	
	\subsubsection{DDR control}
	
	Four read-out modules are connected in a serial connection, and the real-time transmission in 1K frame rate can be achieved when the 5 Gigabit Ethernet is shared.
	On the one hand, in order to avoid data loss caused by the fluctuations, on the other hand, for the sake of achieving higher frame rate transmission in short time. The hardware need a large capacity memories to temporarily store data. This readout electronics design uses 8 GB DDR3 as data cache, while the control of DDR3 is completed by FPGA.
	This firmware is implemented using the Memory Interface Generator (MIG) core and an IP block which supports AXI4 interface provided by Xilinx. By this way, DDR3 can be encapsulated into the form of FIFO interface, which facilitates the application of data cache.

	\subsubsection{TCP packets processing module}

	As the Figure 4 show, the data between the SOP signal and the EOP signal is continuous, there is no gap, and the data is valid under the write signal. User logic can write data from FIFO to XTOE block based on above time sequence. In this design, FIFO is used to handle the processing of cross clock domain and bandwidth mismatch between XTOE and user layer logic. By this way, the timing of XTOE support is encapsulated into FIFO timing, which is very convenient for user layer logic to process. For the user layer, the firmware parallelized the sensor serial signal from the daughter board and wrote data into the 8G DDR3 for caching. The user logic module reads the data from the DDR3 and transfer each  frame into the timing sequence required by the XTOE block. In addition, XTOE core handles data frames into MAC frames and provide an interface with the XGMII bus as the use\cite{10gpcs}.

	\begin{figure}[htbp]
		\centering
		\includegraphics[width=6.47cm]{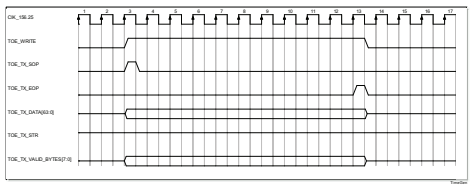}
		\caption{XTOE TCP sequence diagram}
		\label{fig_sim}
	\end{figure}

	\subsubsection{daisy chain connection mode}
	
	To reduce the high complexity of the interfaces and cables, the firmware adopts the connection mode of the daisy chain. It can not only to process the TCP packets from the read-out board, but also the data from the upstream board. Each read-out module is connected to together through Nano-Pitch connectors which with industry-leading port density, multi-protocol support and enhanced signal integrity. Both the upstream and the downstream data are processed simultaneously by the arbitration module with polling operation. This method can enable 4 read-out boards connect together which is equivalent to multiple detector module share the same bandwidth.

	\begin{figure}[htbp]
		\centering
		\includegraphics[width=6.5cm]{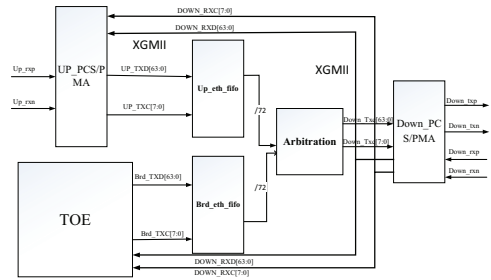}
		\caption{Arbitration module}
		\label{fig_sim}
	\end{figure}

	\begin{figure}[h]
	\centering
	\includegraphics[width=6.5cm]{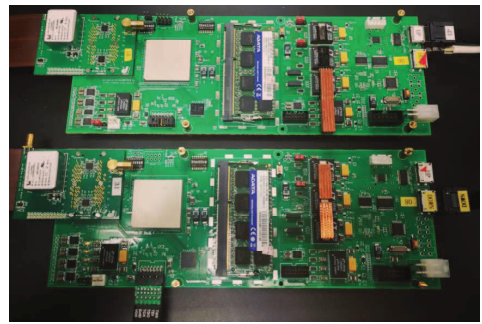}
	\caption{Two read-out boards are connected through Nano-Pitch connector}
	\label{fig_sim}
    \end{figure}
	
	\subsubsection{UDP packets parsing module}

	For UDP parsing module, XTOE block provides the same timing relationship as TCP processing module. The only difference is that they are distinguished by a flag. When dealing with UDP packages, a kind of pattern is adopted to read and write internal\cite{uchida2008hardware}\cite{eiger}. After UDP frames are processed by the XTOE block, the UDP logic module parses the packets and translates them into the addresses and data signals, which can be provided to the user layer to read and write the corresponding register. 
	
	For Sipixel detector, the parameters of the BPIX read-out chip and the setting of the trigger signal required to be configured through the UDP protocol. Because UDP doesn’t like TCP which is a handshake protocol and can guarantee the security of the data, so the firmware provides the read back functionality that can ensure UDP packets successfully configure the register for the whole system.

	\begin{figure}[htbp]
		\centering
		\includegraphics[width=6.5cm]{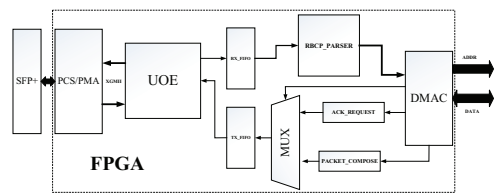}
		\caption{UDP parsing module}
		\label{fig_sim}
	\end{figure}

	\section{Bandwidth measurement}
	
	\subsection{The measurement of XTOE bandwidth}
	
	\subsubsection{Bandwidth measurement between server and XTOE}
	The readout board that instantiated XTOE block as client sends data to the server carrying the CentOS system.
	
	test environment:
	
	\begin{figure}[htbp]
	\centering
	\includegraphics[width=6.5cm]{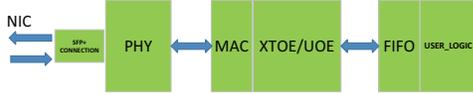}
	\caption{measurement of tcp bandwidth}
	\label{fig_sim}
    \end{figure}

The sources from 0 to 10Gbps are generated via two clock (125MHz 156.25mhz) and different bit width data. As shown in Figure 8, FIFO is used to match data rates of traffic generator and TCP processing module. At the same time, IPERF as the server port measures the throughput .

	linerity test result:

	\begin{figure}[htbp]
		\centering
		\includegraphics[width=6.5cm]{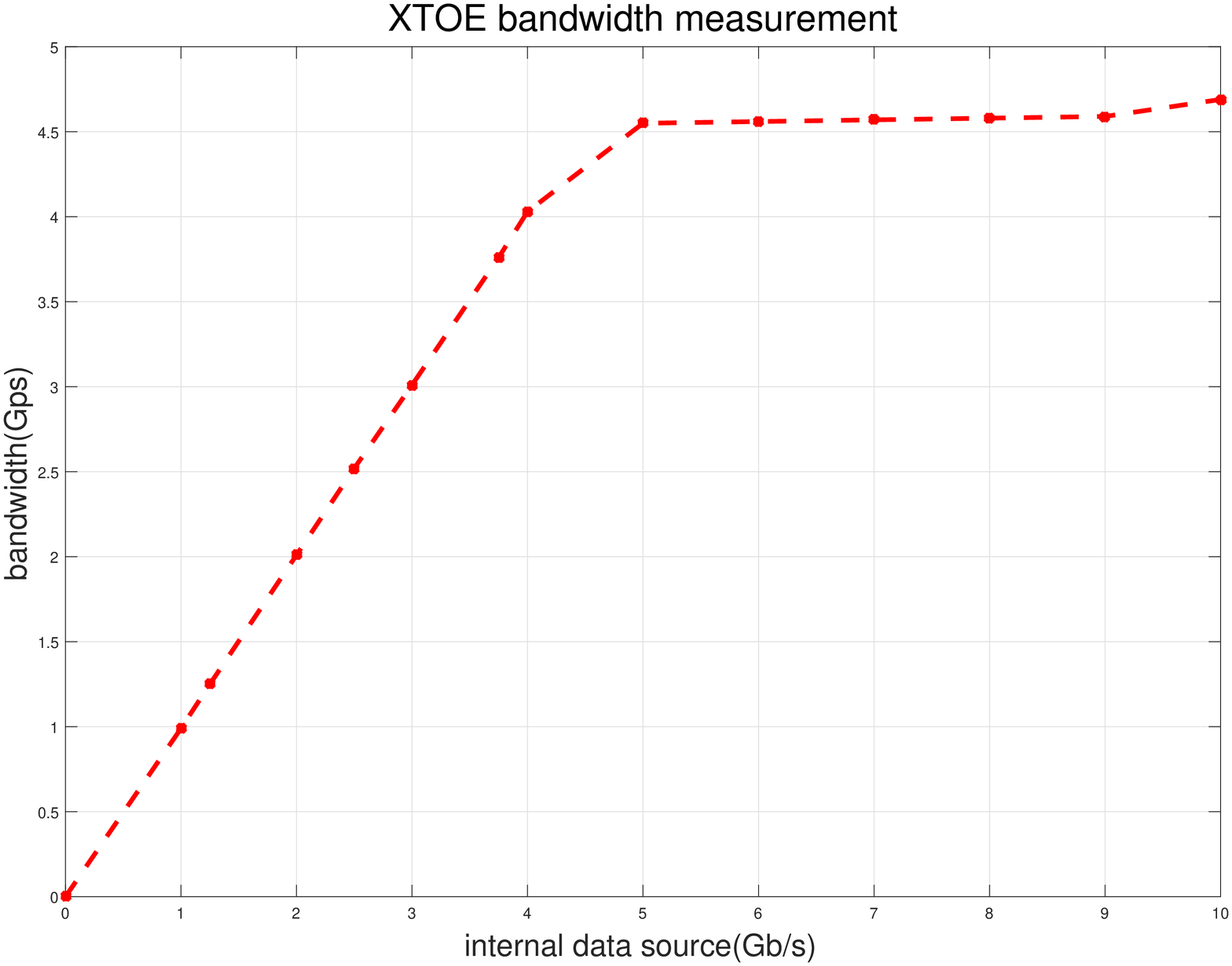}
		\caption{linerity of bandwidth test}
		\label{fig_sim}
	\end{figure}

	\subsubsection{maximum bandwitdh test measurement}

	The measure environment is to use two interconnected readout boards with XTOE IP block, in which FPGA internally generates TCP payload frames. Then it will repeat within the same payload. It also drives the required signals in to the transmit side of FIFO interface. The generator’s data rate is almost 6Gbps (156.25Mhz clock multiplied by 64bit data, but there is a gap between two packet). One of the readout boards receives the data as the server, and the other sends the data as a client. 
	
	\begin{figure}[htbp]
		\centering
		\includegraphics[width=6.5cm]{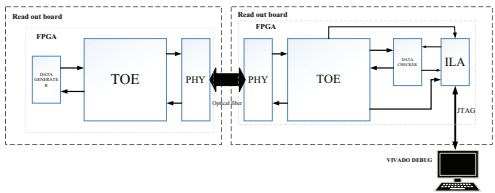}
		\caption{measurement of maximum bandwidth}
		\label{fig_sim}
	\end{figure}

	Two boards are connected each other through the SFP+ module. Vivado debugger  sample the registers’ counts which are calculated by internal logic block and save the data as CSV format, then get the maximum bandwidth by MATLAB. The results are shown in figures 11 and 12.

	\begin{figure}[htbp]
		\centering
		\includegraphics[width=6.6cm]{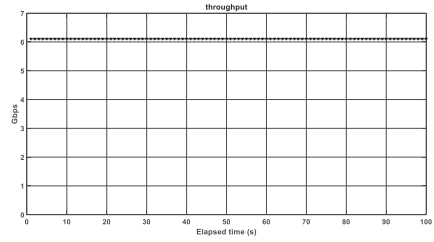}
		\caption{The result of Test throughput between 100 seconds}
		\label{fig_sim}
	\end{figure}
	
The vertical axis shows throughput of TCP payload frames and the horizontal axis shows elapsed time starting from receiving the first data frame. The throughput was recorded by the registers every 100 micro seconds.

	\begin{figure}[htbp]
		\centering
		\includegraphics[width=6.6cm]{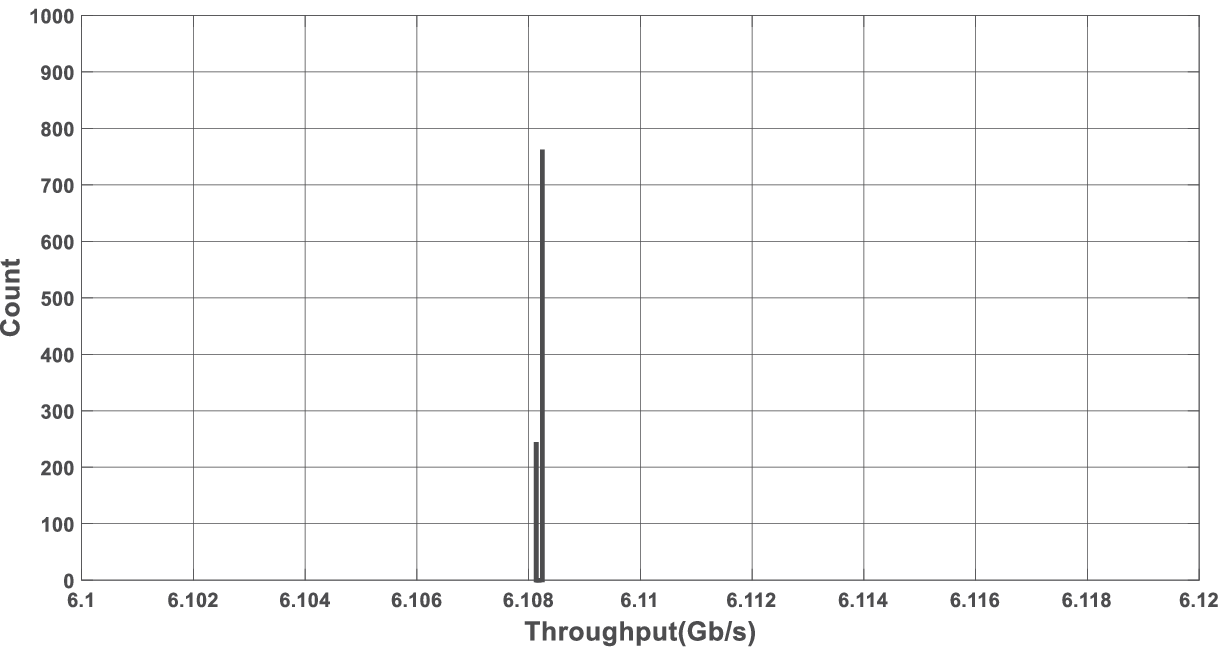}
		\caption{distribution of throughput}
		\label{fig_sim}
	\end{figure}

 Figure 12 shows the distribution of throughput. From the results, the maximum bandwidth of XTOE TCP transmission can be stabilized at 6.108Gbps.

	\subsection{daisy mode measurement}
	In order to test the polling function in the daisy chain mode, each readout board generates data with different bandwidth and connect to each other through Nano-Pitch connectors. Then the last readout board is connected to the server by the SFP+ module and optical cable. With this approach, the bandwidth of total read-out boards can be measured to verify the feasibility of the daisy chain connection mode. The data rate of the readout board is configured through the UDP protocol. After the processing of DAQ, the UDP packets are transmitted and the data packets are translated by the UDP parsing module. The firmware writes registers by this way, so as to achieve the purpose of controlling data rate.

	
	
	

	\section{Conclusion}
	This article introduces a method based on 10 Gigabit FPGA IP block to read out and interact with the silicon pixel detector. The whole hardware is partial verified and tested. Besides, the way of daisy chain mode not only reduces the interface connection of the whole system, but also share the one bandwidth. UDP function has proved that it can read and write registers which are defined by firmware, so that the configuration of the whole detector can be completed.


	\section*{Acknowledgment}

This work was supported by a grant from the National Key Research and Development Program of China (No. 2016YFA0401301)

	
	

\begin{thebibliography}{1}
	
	  
	\bibitem{uchida2008hardware}
	Uchida, Tomohisa, \emph "Hardware-based TCP processor for gigabit ethernet".\hskip 1em plus	0.5em minus 0.4em\relax IEEE Transactions on Nuclear Science, 2008.
	
	\bibitem{gu2017high}
	Gu, Jingzi and Zhang, Jie and Wei, Wei and Ning, Zhe and Li, Zhenjie and Jiang, Xiaoshan and Fan, Lei and Shen, Wei and Ren, Jiayi and Ji, Xiaolu and others, \emph "A High Frame Rate Test System for the HEPS-BPIX based on NI-sbRIO Board".\hskip 1em plus
	0.5em minus 0.4em\relax IEEE Transactions on Nuclear Science, 2017.
	
	\bibitem{eiger}
    Br{\"u}ckner, Martin and Bergamaschi, A and Cartier, S and Dinapoli, R and Fr{\"o}jdh, E and Greiffenberg, D and Johnson, I and Mayilyan, D and Mezza, D and Mozzanica, A and others, \emph "A multiple 10 Gbit Ethernet data transfer system for EIGER".
    
	\bibitem{backplanesieee}
    Backplanes, Over and Cables, Copper, \emph "IEEE Standard for Ethernet".
    
	\bibitem{10gpcs}
    Xilinx, \emph "10G Ethernet PCS/PMA v6.0".
    
	\bibitem{zhang2017heps}
    Zhang, J and Wei, W and Gu, J and Shen, W and Li, Z and Ning, Z and Fan, L and Chen, M and Lu, Y and Ma, X and others, \emph "HEPS-BPIX, a hybrid pixel detector system for the High Energy Photon Source in China".\hskip 1em plus	0.5em minus 0.4em\relax Journal of Instrumentation, 2017.
    
	\bibitem{wei2015high}
    Wei, Wei and Zhang, Jie and Ning, Zhe and Lu, Yunpeng and Fan, Lei and Li, Huaishen and Jiang, Xiaoshan and Lan, Allan K and Ouyang, Qun and Wang, Zheng and others, \emph "A high frame rate pixel readout chip design for synchrotron radiation applications".\hskip 1em plus	0.5em minus 0.4em\relax Nuclear Science Symposium and Medical Imaging Conference (NSS/MIC), 2015 IEEE.
    

	\end{thebibliography}
	%

\end{document}